# Robust Detection of Retinal Neovascularization in Widefield Optical Coherence Tomography

**JINYI HAO,[1] JIE WANG,[1] LIQIN GAO,[1] TRISTAN T. HORMEL,[1] YUKUN GUO,[1,2] AN-LUN WU,[1,3] CHRISTINA J. FLAXEL,[1] STEVEN T. BAILEY,[1] KOTARO TSUBOI,[4] THOMAS S. HWANG,[1] YALI JIA[1,2,*]**

[1]*Casey Eye Institute, Oregon Health & Science University, Portland, Oregon 97239, USA*
[2]*Department of Biomedical Engineering, Oregon Health & Science University, Portland, Oregon 97239, USA*
[3]*Department of Ophthalmology, Mackay Memorial Hospital, Hsinchu 300044, Taiwan*
[4]*Department of Ophthalmology, Aichi Medical University, 1-1, Yazako Karimata, Nagakute, Aichi, 480-1195, Japan*
**jiaya@ohsu.edu*

**Abstract:** Retinal neovascularization (RNV) is a vision threatening development in diabetic retinopathy (DR). Vision loss associated with RNV is preventable with timely intervention, making RNV clinical screening and monitoring a priority. Optical coherence tomography (OCT) angiography (OCTA) provides high-resolution imaging and high-sensitivity detection of RNV lesions. With recent commercial devices introducing widefield OCTA imaging to the clinic, the technology stands to improve early detection of RNV pathology. However, to meet clinical requirements these imaging capabilities must be combined with effective RNV detection and quantification, but existing algorithms for OCTA images are optimized for conventional, i.e. narrow, fields of view. Here, we present a novel approach for RNV diagnosis and staging on widefield OCT/OCTA. Unlike conventional methods dependent on multi-layer retinal segmentation, our model reframes RNV identification as a direct binary localization task. Our fully automated approach was trained and validated on 634 widefield scans (17×17-mm to 26×21-mm) collected from multiple devices at multiple clinics. Our method achieved a device-dependent area under curve (AUC) ranging from 0.96 to 0.99 for RNV diagnosis, and mean intersection over union (IOU) ranging from 0.76 to 0.88 for segmentation. We also demonstrate our method's ability to monitor lesion growth longitudinally. Our results indicate that deep learning-based analysis for widefield OCTA images could offer a valuable means for improving RNV screening and management.

## 1. Introduction

Diabetic retinopathy (DR) is the leading cause of preventable blindness among the working-age population worldwide [1, 2]. In proliferative diabetic retinopathy (PDR), retinal neovascularization (RNV) emerges as a hallmark pathological feature that can lead to severe vision loss due to vitreous hemorrhage or tractional retinal detachment [3]. A key anatomical structure involved in this process is the vitreoretinal interface (VRI), which acts as the boundary between the retina and the vitreous. The penetration of RNV through the VRI marks a critical transition in disease progression, enabling pathological vessels to extend into the vitreous cavity. This disruption significantly increases the risk of vision-threatening complications, underscoring the need for timely clinical intervention to halt progression and preserve vision [4]. But for timely intervention we require highly sensitive screening and monitoring techniques. Fluorescein angiography (FA) is widely used in clinical practice for the assessment of RNV. However, its invasive nature and limited depth resolution motivate the development of complementary, noninvasive approaches [5].

Optimal RNV management requires an imaging modality that can effectively visualize even small RNV lesions, which can occur throughout the retina. Widefield optical coherence tomography (OCT) and OCT angiography (OCTA) enable comprehensive structural and vascular assessment across a large field of view (FOV). Standard OCT reveals retinal structural information such as alterations above the VRI that are associated with RNV [6]. OCTA is generated by measuring motion contrast between OCT cross-sections, which highlights retinal vasculature [7, 8], and can be acquired simultaneously with OCT. Moreover, widefield OCTA is capable of detecting subclinical RNV, and measurements of RNV lesion complexity and area on OCTA have been shown to correlate with treatment response [9]. Finally, both widefield OCT and OCTA are noninvasive, rapid, and cost-effective, making them well-suited for screening.

In addition to an imaging modality capable to accurately identifying RNV, screening and monitoring also require automated image analysis procedures for diagnosis and monitoring. Deep learning has already facilitated automated detection, segmentation, and quantification of RNV and other vascular abnormalities on OCTA images [10-12]. However, these results were achieved on small FOV, a traditional limitation in OCTA imaging that has recently been ameliorated by devices that can obtain much larger FOV images (26 × 21-mm for DREAM OCT, for example, compared to earlier devices that typically obtained 6 × 6-mm FOVs in a single shot at most). RNV detection in widefield OCTA is more challenging. To begin with, retinal anatomy differs from the posterior pole to the periphery, which requires algorithms to be context sensitive. This anatomic variation has several aspects, including thinning of some tissue layers and merging of some plexuses. Of particular importance is the eye's curvature, which introduces axial variation in the scanning region. Furthermore, optical aberrations present in the periphery degrade image quality through blurring (loss of contrast and resolution). These are complications that must be addressed in an effective RNV detection solution for widefield imaging.

In this study, we propose a fully automated deep learning framework that can successfully detect and segment RNV in widefield OCT/OCTA scans. To ensure that the proposed framework can adapt to different RNV morphologies, the vitreous section in the image, defined as the axial region from the detected VRI up to the top of the OCT volume, is designated as the detection zone, rather than relying on a fixed height slab as in previous methods [13, 14]. The framework adopts a two-stage approach: the first stage performs accurate VRI segmentation in order to identify the vitreous part, while the second stage jointly analyzes *en face* OCTA and OCT images to identify and segment RNV lesions. The proposed method is validated on a multi-device OCTA dataset comprising images acquired from two clinics and three imaging systems with varying FOV, demonstrating its generalizability and robustness for fully automated RNV analysis.

## 2. Methods

### *2.1 Data set*

In this study, OCT/OCTA scans were collected using three commercial imaging platforms at two eye institutes, the Casey Eye Institute, Oregon Health & Science University, (Portland, OR, USA); and the Department of Ophthalmology, Aichi Medical University (Nagakute, Japan). The Solix (Visionix/Optovue Inc., Fremont, CA, USA), is a spectral-domain OCT/OCTA platform. It acquires four 9 × 9-mm OCT/OCTA volumes at different retinal locations, collectively covering approximately a 17 × 17-mm region that includes the superotemporal (ST), superonasal (SN), inferotemporal (IT), and inferonasal (IN) regions. Each set is acquired within approximately 5 minutes [15]. The DREAM OCT (Intalight Inc., San Jose, CA, USA) is a swept-source widefield OCTA platform designed for single-shot acquisition. It captures 26 × 21-mm OCTA scans in a single acquisition session lasting around two minutes, including both acquisition and processing time. This approach simplifies imaging for patients with poor fixation and is well suited for clinical workflows, though it involves some degree of down sampling [16]. The S1-OCTA (Canon Inc., Tokyo, Japan) is a swept-source OCTA device operating over a wavelength range of 1010-1160 nm. For each participant, a 23 × 20-mm widefield OCTA image was acquired, centered on the fovea [17]. The acquisition parameters for the three OCT/OCTA systems are summarized (Table 1).

**Table 1. Summary of Acquisition Parameters for the Three OCT/OCTA Systems**

| Device | FOV (mm) | A-scans × B-frames* | A-scan rate (kHz) | Sampling density ($\mu m$/pixel) |
|---|---|---|---|---|
| Solix | 9 × 9** | 600 × 600 | 120 | 15 × 15 |
| DREAM OCT | 26 × 21 | 1512 × 1240 | 200 | 17 × 17 |
| S1-OCTA | 23 × 20 | 928 × 807 | 100 | 25 × 25 |

*B-frame: A cross-sectional frame generated by processing multiple repeated B-scans at the single location to capture motion contrast.
**For the Solix system, widefield retinal coverage (~17 × 17-mm) was achieved by acquiring four 9 × 9-mm scans at different retinal locations.

The dataset comprised both RNV-positive and non-RNV control cases. All RNV cases were clinically confirmed by retinal specialists based on multimodal imaging and clinical diagnosis. The non-RNV control group consisted of both healthy eyes and eyes diagnosed with other retinal conditions, including age-related macular degeneration (AMD), non-neovascular diabetic retinopathy, and branch retinal vein or artery occlusion (BRVO/BRAO), each confirmed by clinical examination. For non-RNV controls only one eye was imaged per volunteer. For RNV cases, both eyes were included if both presented with RNV, otherwise only the eye with RNV was used in the dataset. Furthermore, for patients with longitudinal imaging records, each follow-up scan was treated as an independent case due to morphological changes between visits. No scans were excluded due to poor image quality.

### 2.2 Preprocessing

To address domain shifts caused by variations in OCTA imaging devices, all input images were subjected to a standardized preprocessing pipeline. First, column-wise z-score normalization was applied, whereby each column of the image was normalized using its own mean and standard deviation [18]. This technique reduces directional intensity variation and enhances inter-device consistency. For multi-channel inputs (OCT and OCTA), this normalization was performed independently on each channel to preserve modality-specific contrast characteristics.

To further suppress background noise and enhance lesion visibility, a thresholding operation was applied following normalization. Specifically, pixel values below -0.5 were set to zero, effectively attenuating low-intensity artifacts and improving the signal-to-background ratio in regions affected by noise or motion-induced fluctuations.

### 2.3 Ground truth label generation

In this study, two types of ground truth labels were generated, each designed for a specific model task: VRI segmentation, and RNV diagnosis with membrane segmentation.

In this study, retinal specialists manually annotated the VRI boundary on each B-scan. Using these annotations, binary segmentation masks were generated by labeling pixels above the VRI as vitreous and those below as the retinal region.

*En face* projections of the vitreous slab were used to annotate RNV lesions. Based on these projections, retinal specialists manually annotated RNV as follows: first, hyperreflective structures located on *en face* OCT of the vitreous slab were identified as candidate lesions [19]; second, their boundaries were carefully delineated on *en face* OCTA images. The annotations generated from the *en face* OCTA image, where features such as tangled vasculature or abnormal sprouting patterns were identified, and subsequently verified on cross-sectional OCT overlaid by OCTA (Fig. 1A). Each annotation was independently reviewed and confirmed by at least two retinal specialists (Fig. 1E).

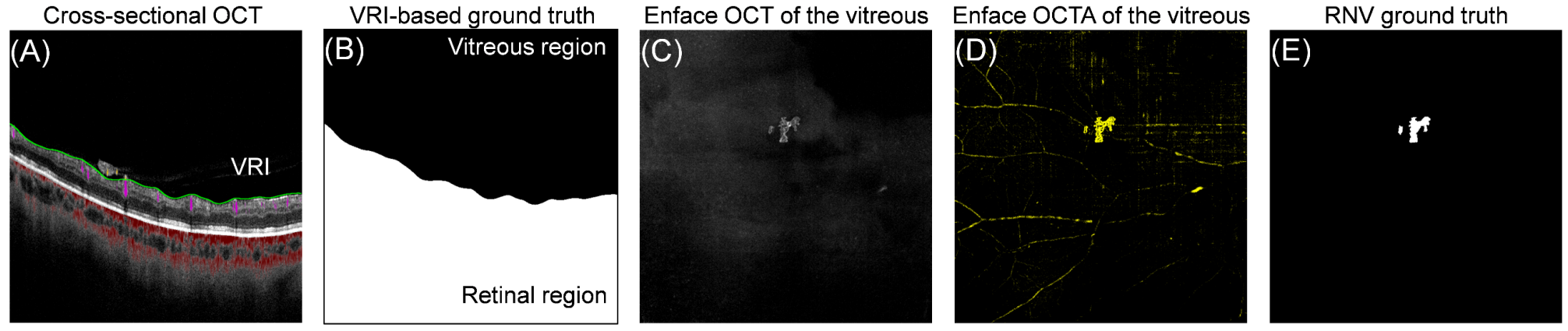

**Fig. 1.** Ground truth label generation workflow. (A) Cross-sectional OCT image showing expert-annotated VRI boundary with flow signal overlaid (yellow: RNV flow; violet: retinal flow; red: choroidal flow). (B) Binary VRI-based ground truth mask showing the vitreous (black) and retina (white). (C) *En face* OCT projection of the vitreous slab, used to identify candidate hyperreflective structures suggestive of RNV. (D) *En face* OCTA projection of the vitreous slab, used to delineate the final RNV boundaries. (E) Manually annotated RNV ground truth, The lesion locations were first identified on *en face* OCT images, and the membrane boundaries were delineated on *en face* OCTA based on the clearer flow-signal contrast between RNV and background. The enclosed regions were then filled to generate the final ground-truth mask.

### 2.4 Algorithm outline

The proposed algorithm incorporates two convolutional neural networks (CNNs): one for VRI segmentation and the other for *en face* RNV detection (Fig. 2). The VRI segmentation module serves as the foundation for subsequent RNV analysis by generating anatomically consistent *en face* OCTA and OCT projections.

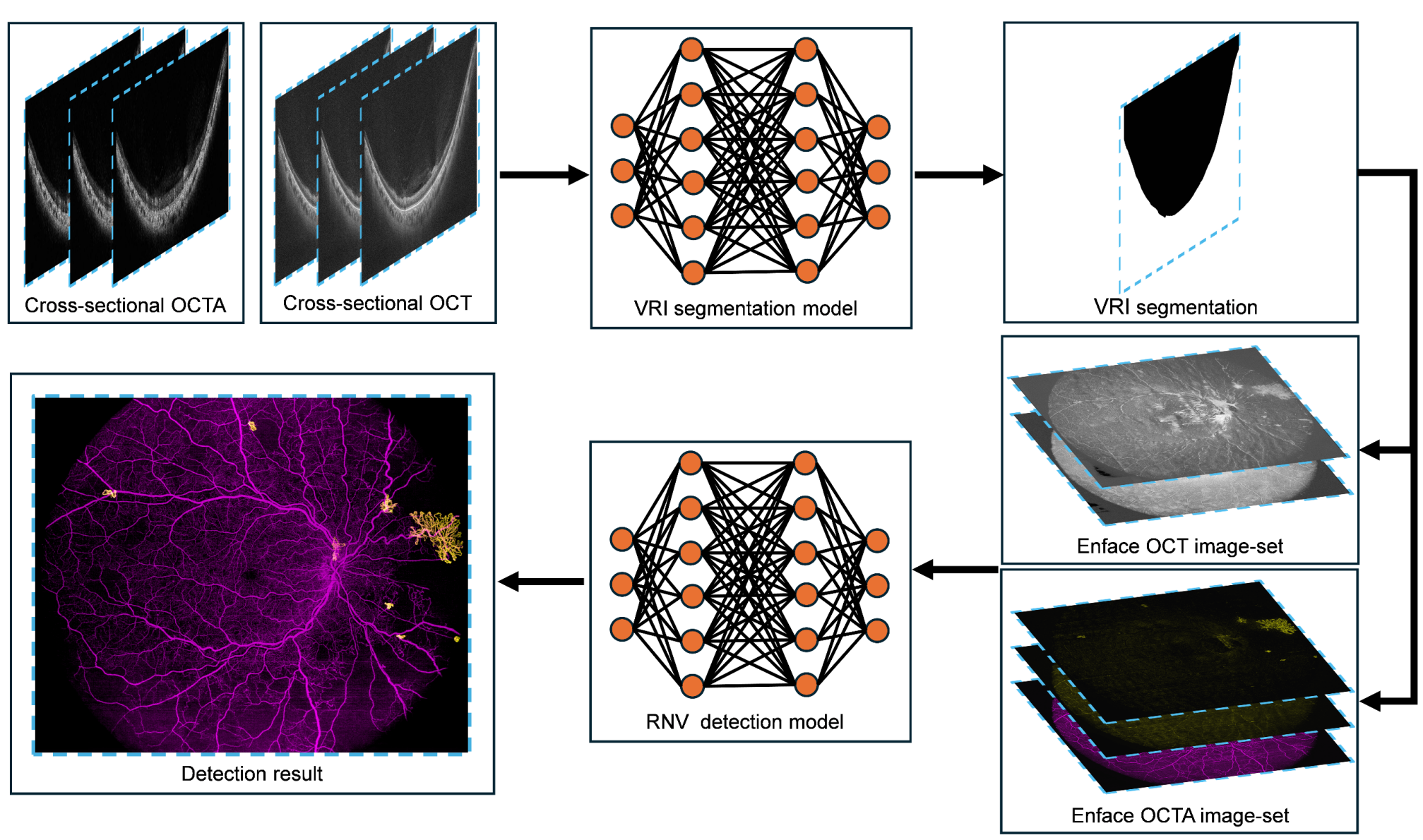

**Fig. 2.** Schematic overview of the proposed automated RNV identification and segmentation framework. The input consists of cross-sectional structural OCT and OCTA images, while the final output is a pixel-wise RNV segmentation map visualized on the *en face* OCTA projection of the vitreous slab onto the corresponding *en face* OCTA of the inner slab. Two separate CNNs are employed: the first segments the VRI boundary, which is then used to generate *en face* OCT and OCTA image sets; the second CNN detects and segments the RNV membrane based on these anatomically guided *en face* OCT and OCTA inputs.

Based on the predicted VRI, multiple anatomically relevant *en face* projection slabs are generated from both OCT and OCTA volumes. These projection slabs are subsequently fed into the RNV detection network. When RNV is present, the model produces a pixel-wise segmentation mask outlining the lesion area. The resulting detection map represented the predicted lesion region and was used for subsequent quantitative overlap and detection analyses. For visualization purposes, the flow signals confined to the predicted lesion area were extracted from *en face* OCTA of the vitreous slab and projected onto the corresponding *en face* OCTA of the inner slab. Otherwise, scans without detectable lesions are classified as non-RNV.

### *2.5 Deep learning model for VRI segmentation*

To segment the VRI, we designed a deep learning model based on the U-Net architecture [20]. Although the VRI may appear smooth in certain cases, its localization becomes substantially more challenging in the presence of RNV, particularly for flat-type lesions that are tightly adherent to the retinal surface. In such cases, reliable localization requires contextual information from adjacent B-scans. Motivated by this challenge, the proposed network takes as input three consecutive B-scan frames, with each frame comprising paired structural OCT and OCTA images. These six images are concatenated along the channel axis to form a 6-channel input tensor, enabling the model to leverage local volumetric context. This multi-frame, dual-modality configuration provides complementary structural and vascular information, which is useful for robustly identifying the vitreous region and minimizing segmentation errors caused by noise or morphological variability (Fig. 3). The network outputs a binary segmentation mask, classifying each pixel as either in the vitreous region above the VRI or the retinal region below the VRI.

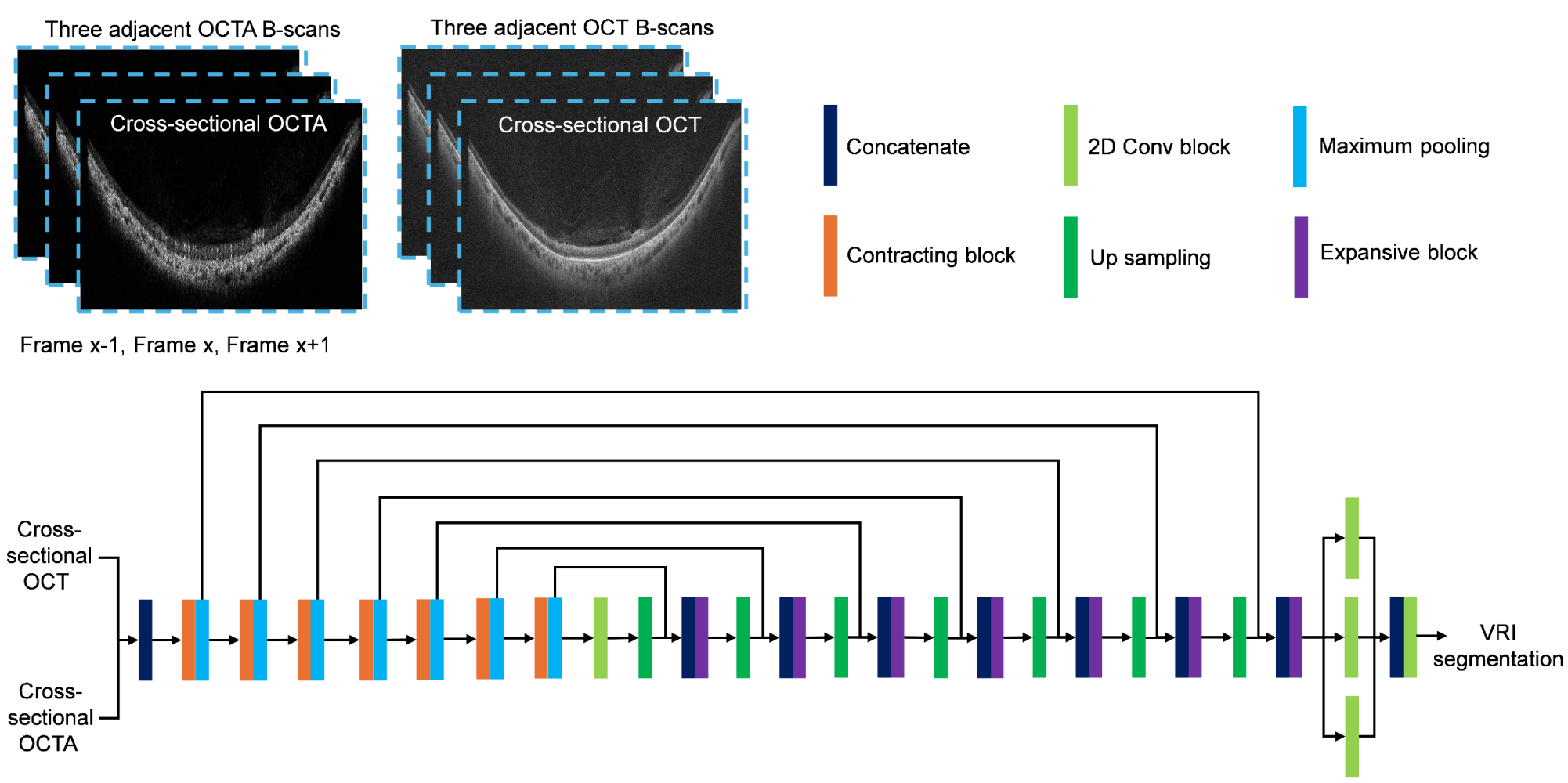

**Fig. 3.** Schematic CNN architecture for VRI boundary segmentation. The model inputs three adjacent OCT and OCTA B-scans to capture both structural and vascular features with local 3D context. Multi-scale feature fusion is applied using parallel 1 × 1, 3 × 3, and 5 × 5 convolutions, enabling accurate and robust VRI segmentation across varied retinal structures.

To optimize the VRI segmentation model we adopted a composite loss function that combines Binary Cross-Entropy (BCE) Loss [21] and Dice Loss, two widely used metrics in medical image segmentation tasks. BCE Loss ensures pixel-wise classification accuracy by penalizing incorrect predictions at the individual pixel level. In contrast, Dice Loss measures the spatial overlap between the predicted and ground truth regions, making it particularly effective for addressing class imbalance, which is common in anatomical boundary segmentation. The final loss function is formulated as a weighted sum of the two components to leverage their complementary strengths and achieve more robust and accurate VRI delineation:

$$\mathcal{L}_{BCE} = -\frac{1}{N}\sum_{i=1}^{N}\left(y_i \log(\hat{y}_i) + (1 - y_i)\log(1 - \hat{y}_i)\right) \tag{1}$$

$$\mathcal{L}_{Dice} = 1 - \frac{2\sum_{i=1}^{N} y_i \hat{y}_i + \epsilon}{\sum_{i=1}^{N} y_i + \sum_{i=1}^{N} \hat{y}_i + \epsilon} \tag{2}$$

$$\mathcal{L}_S = \alpha \mathcal{L}_{BCE} + (1-\alpha)\mathcal{L}_{Dice} \tag{3}$$

Here, $y_i$ denotes the ground truth label of the $i$-th sample, and $\hat{y}_i$ represents the corresponding predicted probability. $N$ is the total number of samples in the dataset. A small constant $\epsilon$ is added to both the numerator and denominator to avoid division by zero and ensure numerical stability during training. $\mathcal{L}_S$ is defined as the final loss value used for training the VRI segmentation network, and $\alpha$ is a weighting factor that balances the contributions of the BCE and Dice loss components, with its numerical value set to 0.5 in our experiments.

### *2.6 Deep learning model for RNV diagnosis and segmentation*

The next step, RNV segmentation, uses *en face* projections generated from both OCT and OCTA volumes based on the VRI segmentation using the maximum projection method [22]. By isolating the vitreous, this approach reduces segmentation artifacts from superficial vessels and suppresses confounding signals. Specifically, five types of *en face* images were constructed as model inputs, each capturing complementary structural or vascular cues relevant to neovascular pathology and supporting accurate RNV diagnosis and membrane segmentation (Fig. 4).

**(1) *en face* OCT of the vitreous**

The vitreous slab was defined using the VRI as the lower boundary and extended axially upward without a fixed height. This definition enables comprehensive coverage of neovascular structures in the vitreous that vary substantially in height and morphology. *En face* OCT projection of this slab highlights hyperreflective lesions in the vitreous that are strongly indicative of pathological neovascular proliferation.

**(2) *en face* OCTA of the vitreous**

The *en face* OCTA projection of the same slab provides a detailed visualization of abnormal blood flow within the vitreous region, which is avascular in normal eyes.

**(3) *en face* OCT of GCC**

The ganglion cell complex (GCC) slab was defined with the VRI as the upper boundary and extended 80 μm into the retina, corresponding to the typical thickness of the ganglion cell complex [23]. In advanced stages, RNV can exert mechanical traction or pressure on the inner retinal layers, resulting in localized compression or deformation that can offer important complementary useful for RNV detection [24].

**(4) *en face* OCTA of GCC**

The *en face* OCTA projection of the GCC slab provides spatial priors of large superficial retinal vessels that can protrude slightly towards the VRI and elevate its contour. Combined with minor inaccuracies in VRI segmentation, this can produce segmentation artifacts appearing as flow in the vitreous. Incorporating the GCC-layer OCTA into the model input enables the network to recognize and discount such artifacts based on their spatial distribution patterns.

**(5) subtracted *en face* OCTA**

The subtracted *en face* OCTA image of the vitreous was generated by subtracting a scaled version of the GCC-layer OCTA signal from the vitreous OCTA image. This subtraction enhances the contrast of neovascular tufts by reducing artifacts from superficial vasculature caused by segmentation inaccuracies, as demonstrated in previous work on OCTA image enhancement [25, 26].

To preserve modality-specific features, a dual-branch encoding design was adopted, allowing OCT and OCTA inputs to be processed separately during early feature extraction.

To enhance detection sensitivity and structural fidelity, several architectural modifications were integrated into the U-Net backbone. Squeeze-and-excitation (SE) blocks are inserted after each encoder and decoder stage to recalibrate channel responses [27], improving the model's ability to detect small or low-contrast neovascular tufts. In addition, depthwise separable convolutions replace standard convolutions in the second layer of each block [28], reducing computational complexity while preserving discriminative power. Multi-scale fusion and skip connections are used to support both high-resolution localization and contextual understanding.

To address the inherent class imbalance in RNV segmentation, where the neovascular regions (foreground) occupy only a small portion of the image compared to the large background, we adopted a loss function that explicitly separates and balances the contributions of both foreground and background. Specifically, we compute Mean Squared Error (MSE) losses independently for the foreground and background, then take their average as the final loss:

$$\mathcal{L}_D = \frac{\omega_b}{N_b + \epsilon} \sum_{i}^{\mathbb{I}_{b,i}} (y_i - \hat{y}_i)^2 + \frac{\omega_f}{N_f + \epsilon} \sum_{i}^{\mathbb{I}_{f,i}} (y_i - \hat{y}_i)^2 \tag{4}$$

$\mathcal{L}_D$ is defined as the final loss value used for training the RNV membrane segmentation network. In this formulation, $\mathbb{I}_{b,i}$ and $\mathbb{I}_{f,i}$ are indicator functions that select background and foreground pixels respectively; $N_b$ and $N_f$ the total number of background and foreground pixels; and $\epsilon$ is a small constant to avoid division by zero, and $\omega_b$, $\omega_f$ are weighting factors for background and foreground losses, respectively, set to 0.4 and 0.6. This formulation helps the model better capture neovascular regions despite their small size in the image.

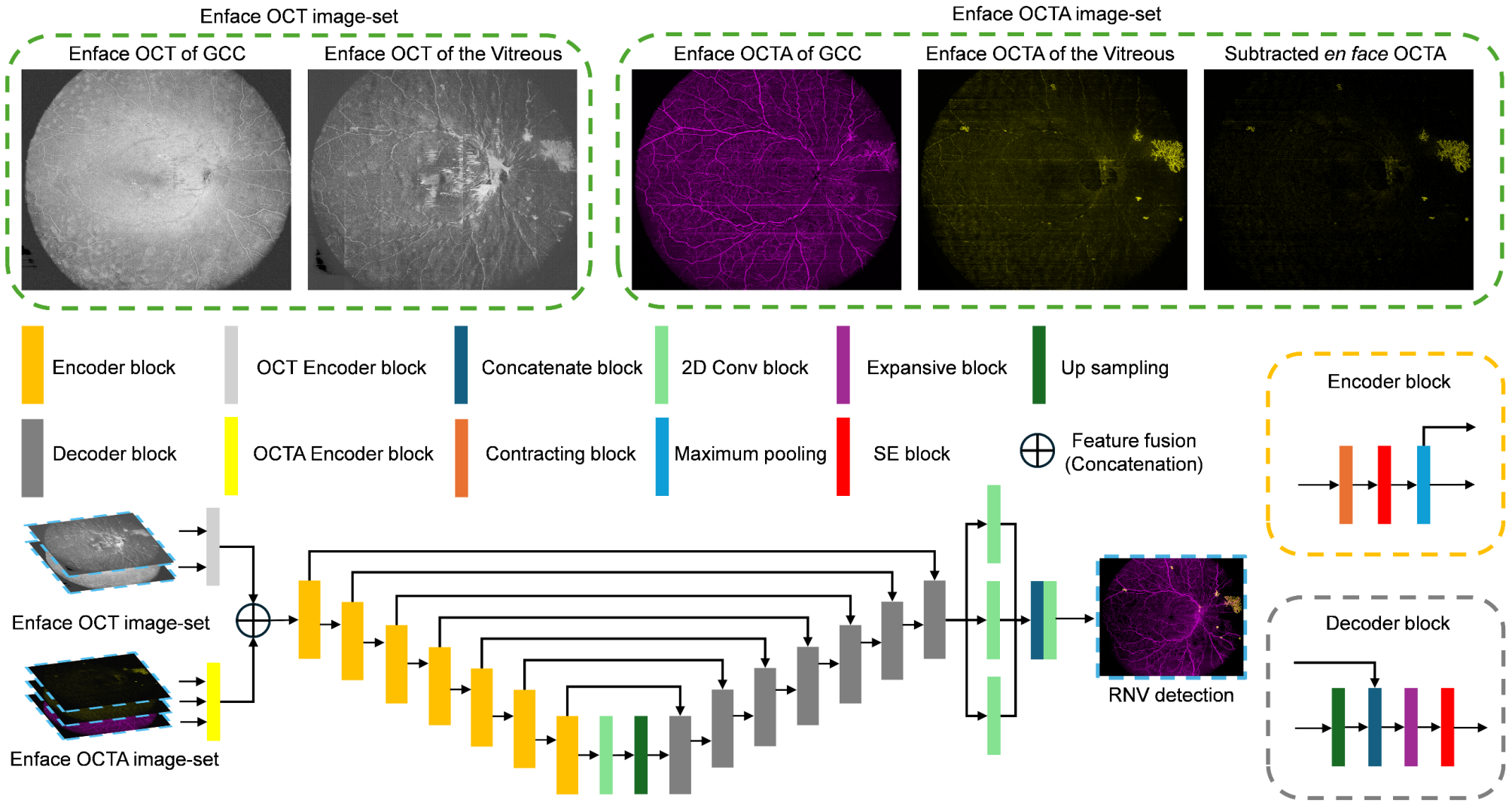

**Fig. 4.** Overview of the dual-branch CNN for RNV membrane detection. *En face* OCT and OCTA image sets are processed through independent encoder branches. The OCT and OCTA encoder branches share the same network architecture but use independent weights and are trained separately to extract features. These features are then fused via concatenation and fed into the first encoder layer of a U-Net backbone. Each encoder block integrates standard and depthwise convolutions, squeeze-and-excitation (SE) attention, and pooling. Decoder blocks include up sampling, skip connections, and SE recalibration. A multi-scale fusion head generates the final segmentation.

## *2.7 Training and evaluation*

To enhance the robustness and generalization ability of the model, data augmentation was applied to 50% of the samples in each training batch. Two augmentation strategies were employed: (1) brightness adjustment, in which image intensities were randomly scaled to simulate varying illumination conditions; and (2) random Gaussian noise (mean = 0, standard deviation = 0.1) was added to the input images to improve robustness to noise. These augmentations were applied to both structural OCT and OCTA inputs.

To ensure unbiased evaluation, scans from the same eye were never included in both training and testing sets. Under this eye-level separation constraint, the data were split such that the numbers of scans in the training and testing sets were approximately in a 7:3 ratio. Due to the limited number of eyes in each dataset, an exact 7:3 split at the scan level is not always feasible, resulting in deviations from the nominal ratio. The network was implemented in PyTorch and trained on an NVIDIA RTX 4090 GPU with CUDA 11.8. We optimized the model using the Adam optimizer with a learning rate of $1 \times 10^{-4}$ and applied early stopping based on validation Dice loss with a patience of 20 epochs to prevent overfitting. Training was conducted for a maximum of 1000 epochs with a batch size of 2, and model weights were initialized with a ReLU-specific variance scaling method for stable training [29].

## 3. Results

### 3.1 Dataset composition

In total, 476 scans (including repeat and follow-up sessions) were collected from the Solix device across different retinal regions. An additional 130 scans were collected from the DREAM OCT device, and 28 independent scans without repeat imaging were collected from the S1-OCTA device (Table 2).

**Table 2. Distribution of RNV and Non-RNV Cases Across Devices and Scan Sizes**

| Device | Scan Size (mm) | Training (Eyes/Scans) | | Testing (Eyes/Scans) | |
|---|---|---|---|---|---|
| | | RNV | Non-RNV | RNV | Non-RNV |
| Solix | 9 × 9* | 30/168 | 31/184 | 18/50 | 21/74 |
| DREAM OCT | 26 × 21 | 43/56 | 29/48 | 23/26 | 12/14 |
| S1-OCTA | 23 × 20 | 18/18 | 1/1 | 8/8 | 1/1 |

*For the Solix system, widefield retinal coverage (~17 × 17-mm) was achieved by acquiring four 9 × 9-mm scans at different retinal locations; individual 9 × 9-mm scans were used for training and evaluation, and no scans from the same eye were split across training and testing sets.

### 3.2 VRI segmentation

To evaluate VRI segmentation accuracy, we selected B-scans containing RNV regions from each subject. Two graders independently delineated the VRI boundary using our customized COOL-ART software [30] and the final reference standard was obtained by averaging their annotations. The absolute error of the model-predicted boundary was then computed with respect to this reference, reported as mean ± standard deviation (in pixels). Notably, the segmentation error was calculated over lesion-centered regions, including an expansion of several pixels into adjacent normal tissue instead of the entire scan. This pixel-wise boundary error analysis provides a direct and anatomically meaningful assessment of segmentation performance.

To examine the contribution of OCTA to VRI segmentation, we performed an ablation analysis on the DREAM OCT dataset, comparing structural OCT-only and combined OCT-OCTA inputs (Table 3). Across all test cases, incorporating OCTA resulted in a lower mean segmentation error, however, the difference was not significant (paired Wilcoxon signed-rank test, $\alpha = 0.05$). In contrast, a significant reduction in segmentation error was observed in flat-type RNV cases, suggesting improved segmentation performance in these challenging cases.

**Table 3. Ablation Study of VRI Segmentation on the DREAM OCT Dataset**

| Dataset | Data Size (eyes / scans) | Input Modality | Segmentation Error * (Mean ± std, pixels) |
|---|---|---|---|
| All test cases | 35 / 40 | OCT only | 1.91 ± 3.86 |
| | | OCT + OCTA | 1.32 ± 3.15 |
| Flat-type RNV | 11 / 11 | OCT only | 5.95 ± 3.66 |
| | | OCT + OCTA | 2.28 ± 3.56 |

*The difference was insignificant across all cases ($p \approx 0.8$) but was significant in flat-type RNV cases ($p = 0.002$).

We also report the inter-grader variability across different devices, which quantifies the difference between manual VRI delineations from the two graders. The model segmentation error falls within the range of inter-grader variability across devices (Table 4).

**Table 4. Performance of VRI Segmentation on Different Devices**

| Device | Axial Sampling Resolution ($\mu m$/pixel) | Segmentation Error (Mean ± std, pixels) | Inter-grader Variability (Mean ± std, pixels) |
|---|---|---|---|
| Solix | 3.05 | 0.65 ± 1.51 | 1.63 ± 1.97 |
| DREAM OCT | 2.00 | 1.32 ± 3.15 | 1.81 ± 2.05 |
| S1-OCTA | 4.00 | 2.13 ± 4.77 | 2.20 ± 2.72 |

In addition to the quantitative evaluation reported in Table 4, we further assess the qualitative performance of our VRI segmentation algorithm using three representative OCTA cases exhibiting distinct RNV morphologies. RNV is commonly presented in three morphological types: tabletop, forward, and flat. Flat-type RNV lies tightly adherent to the VRI, making it difficult to distinguish the boundary between pathological and normal tissue (Fig. 5 Row 1). The other two types separate further from the VRI, making the segmentation task easier, although the algorithm still must learn the features associated with each. Our approach is capable of dealing with each type (Fig. 5).

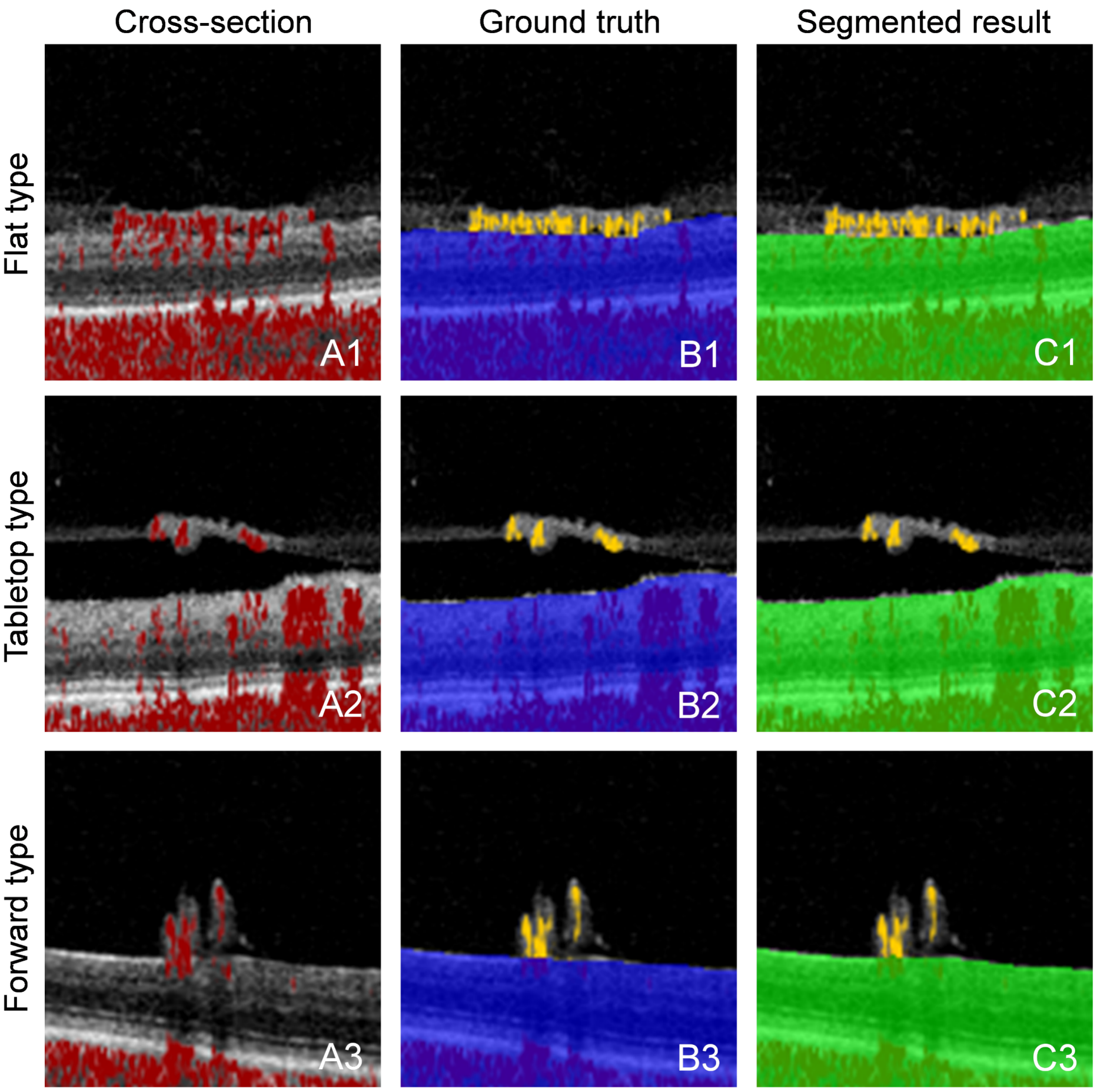

**Fig. 5.** Qualitative evaluation of VRI segmentation performance across three representative RNV morphologies: flat type (Row 1), tabletop type (Row 2), and forward type (Row 3). For visualization purposes, all images are cropped to the region containing the RNV. (A) Original cross-sectional OCT image acquired using the DREAM OCT system with all flow signals shown in red (prior to VRI segmentation). (B) Manually annotated ground truth with flow signal color-coded according to VRI position, red for normal retinal vasculature (including inner retinal and choroidal vessels), and yellow for RNV located in the vitreous; the retinal region below the VRI is shown in blue. (C) Segmented result, where the retinal region below the VRI is shown in green, while the color coding of flow signals is kept consistent with the ground truth. In each case, although the RNV morphologies varied substantially, the VRI segmentation agreed with grader annotations.

The VRI is often smooth and can be localized using simple image-processing strategies in small-FOV OCT, whereas approaches that rely on explicit boundary detection become unreliable due to increased retinal curvature and anatomical variability in widefield OCT. More practically for widefield OCT, we considered a segmentation approach based on relative reflectance, which is referred to as the conventional approach in the following analysis. In this approach, Bruch's membrane (BM) was first localized based on its characteristic high-intensity signal. Using intensity thresholding, the vitreous region including RNV membrane (weak reflectance) can be separated from the retinal slab (strong reflectance). In the presence of RNV, relatively hyporeflective gaps between the neovascular tissue and the inner retinal slab were incorporated to guide VRI localization. While this reflectance-based approach is intuitive and easy to implement, it becomes unreliable in challenging cases such as flat-type RNV, where the neovascular membrane is tightly adherent to the retinal surface and exhibits limited contrast within a single B-scan (Fig. 6). We speculated that accurate segmentation in such cases would require contextual information from adjacent B-scans, which motivated the proposed network design that takes as input three consecutive B-scan frames (both the reflectance and flow signal from each B-frame are included in our dual encoder strategy). This proposed method achieved substantially lower segmentation error than a conventional approach in flat type RNV cases (n=11; Table 5). The impact of segmentation accuracy is most evident when we analyze the location of RNV membranes. If the VRI is incorrectly localized below the RNV membranes, those lesions are excluded from the vitreous slab, resulting in missing or incomplete RNV visualization in the *en face* images.

**Table 5. Comparison of VRI Segmentation Error Between Conventional and Proposed Methods in Flat-Type Cases (n =11)**

| Method | Segmentation Error (Mean ± std, pixels) |
|---|---|
| Conventional approach | 9.74 ± 8.06 |
| Proposed U-Net–based method | 2.28 ± 3.56 |

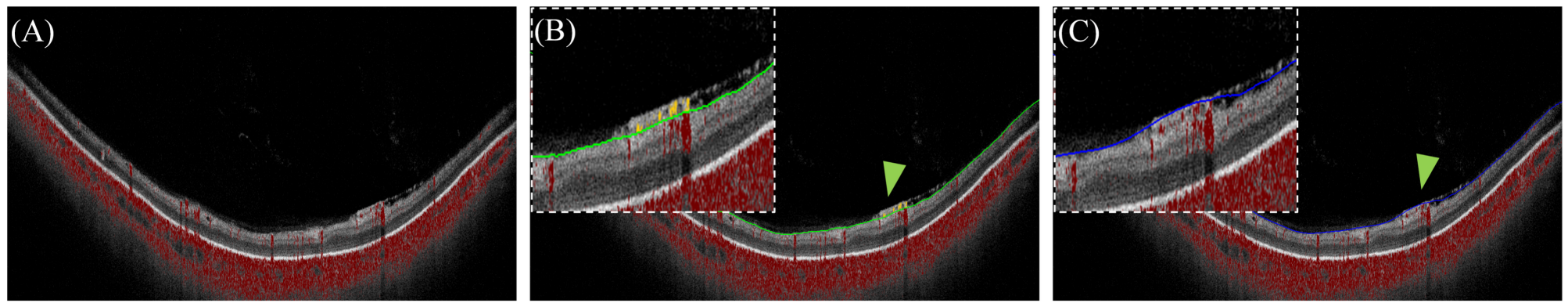

**Fig. 6.** Qualitative evaluation of VRI segmentation in a representative flat-type RNV case. (A) Original cross-sectional OCT image acquired using the DREAM OCT system with flow signal shown in red (prior to VRI segmentation). (B) Segmentation results from the proposed method with flow signals color-coded according to VRI position, red for normal retinal vasculature (including inner retinal and choroidal vessels), and yellow for RNV located in the vitreous. The green line represents VRI. (C) Segmentation results obtained using conventional approach. The blue line represents VRI. Here, no vessels were found in the vitreous according the VRI segmentation result. Magnified views highlight the RNV region.

### *3.3 RNV diagnosis and segmentation*

#### 3.3.1 Quantitative evaluation

The performance of RNV diagnosis was evaluated at the diagnostic level, where the presence of any neovascular lesion within an eye was considered a positive diagnosis. The model was assessed based on its ability to correctly identify the presence of each lesion. For datasets such as Solix, eye-level diagnosis was derived by aggregating the results from all scans acquired for the same eye. In contrast, segmentation performance was evaluated at the pixel level by comparing the predicted lesion masks to the manual annotations using IOU and F1 Score.

$$\text{Precision} = \frac{\text{TP}}{\text{TP} + \text{FP}} \tag{5}$$

$$\text{Recall} = \frac{\text{TP}}{\text{TP} + \text{FN}} \tag{6}$$

$$\text{F1} = 2 \times \frac{\text{Precision} \times \text{Recall}}{\text{Precision} + \text{Recall}} \tag{7}$$

$$\text{IOU} = \frac{\text{GT} \cap \text{Out}}{\text{GT} \cup \text{Out}} \tag{8}$$

Here proposed algorithm. TP, FP, and FN refer to the numbers of true positives, false positives, and false negatives, respectively. The proposed algorithm achieved high performance in both diagnostic (AUC and sensitivity) and segmentation (IOU and F1-score) tasks (Table 6).

**Table 6. Performance of RNV Diagnosis and Membrane Segmentation on Different Devices**

| | RNV Diagnosis | | RNV Segmentation | |
|---|---|---|---|---|
| **Device** | **AUC** | **Sensitivity** | **F1 Score** | **IOU** |
| Solix | 0.99 | 0.92 | 0.92±0.20 | 0.88±0.22 |
| DREAM OCT | 0.98 | 0.85 | 0.90±0.25 | 0.87±0.27 |
| S1-OCTA | 0.96 | 0.75 | 0.79±0.35 | 0.76±0.37 |

We further evaluated the proposed method in comparison with representative segmentation baselines, including SegNet [31], AUNet [32], and BiSeNet [33] on the DREAM OCT dataset. Under the same training and evaluation protocol, the proposed approach achieved comparable or better performance in both RNV diagnosis and membrane segmentation (Table 7).

**Table 7. Performance of RNV Diagnosis and Membrane Segmentation for Different Segmentation Baselines**

| | RNV Diagnosis | | RNV Segmentation | | | |
|---|---|---|---|---|---|---|
| **methods** | **AUC** | **Sensitivity** | **F1 Score** | **p (F1) *** | **IOU** | **p (IOU) *** |
| SegNet | 0.96 | 0.81 | 0.85±0.31 | <0.01 | 0.82±0.33 | <0.01 |
| AUNet | 0.98 | 0.85 | 0.86±0.30 | <0.01 | 0.84±0.32 | <0.01 |
| BiSeNet | 0.94 | 0.77 | 0.73±0.40 | <0.01 | 0.70±0.42 | <0.01 |
| Ours | 0.98 | 0.85 | 0.90±0.25 | — | 0.87±0.27 | — |

*p-values were obtained using paired Wilcoxon signed-rank tests on eye-level aggregated pixel-level F1 and IOU metrics, comparing the proposed method with each baseline.

A comparison with representative RNV-related methods reported in the literature is also provided (Table 8). These approaches focused on small-field OCTA scans and perform eye-level detection, region-of-interest (ROI)-level localization, or B-scan-based neovascularization of the optic disc (NVD) segmentation with 3D reconstruction, whereas the proposed method targets pixel-level RNV membrane segmentation on widefield OCT/OCTA data.

**Table 8. Summary of Representative RNV-Related Methods Reported in the Literature**

| **Reference** | **Modality** | **Task Level** | **Dataset** | **Device Number** | **Data Size (scans)** | **FOV (mm)** | **Reported Performance** |
|---|---|---|---|---|---|---|---|
| Li et al. [10] | OCTA (*en face*) | DR diagnosis and coarse lesion analysis | Challenge dataset (DRAC22) | 1 | 109 | 12×12 | Not reported (task-specific challenge metrics) |
| Tun et al. [12] | OCTA (*en face*) | ROI-level RNV localization | Public (TUH / DRAC22) | 2 | 160 | 12×12 | Precision: 0.90<br>F1: 0.85 |
| Wang et al. [34] | OCT / OCTA | B-scan-based NVD segmentation with 3D reconstruction | Private | 1 | 156 | 6×6 | IOU: 0.79±0.33<br>Precision: 0.86±0.06 |
| Ours | OCT / OCTA | Pixel-level RNV membrane segmentation | Private | 3 | 634 | 9 × 9*<br>26 × 21<br>23 × 20 | IOU: 0.87±0.27<br>F1: 0.90±0.25 |

*For four 9 × 9-mm scans covering the superotemporal (ST), superonasal (SN), inferotemporal (IT), and inferonasal (IN) regions.

### 3.3.2 Performance demonstrated on three devices

To further illustrate its robustness across different imaging conditions, we highlight representative examples from the three OCTA devices. The Solix device provides the highest image resolution among the three imaging systems, enabling clear visualization of *en face* OCTA projections (Fig. 7 row 1). Although prominent superficial vessel signals are visible in the vitreous, the proposed algorithm accurately distinguishes true RNV lesions from these overlying vessels. The DREAM OCT device, with its ultra-wide scanning area of 26 × 21-mm, enables visualization of large-scale RNV lesions (Fig. 7 row 2). Despite the extensive membrane area in this example, the algorithm successfully captures the full extent of the lesion, allowing for reliable computation of lesion-level quantitative metrics. The S1-OCTA device also offers a wide FOV (Fig. 7 row 3).

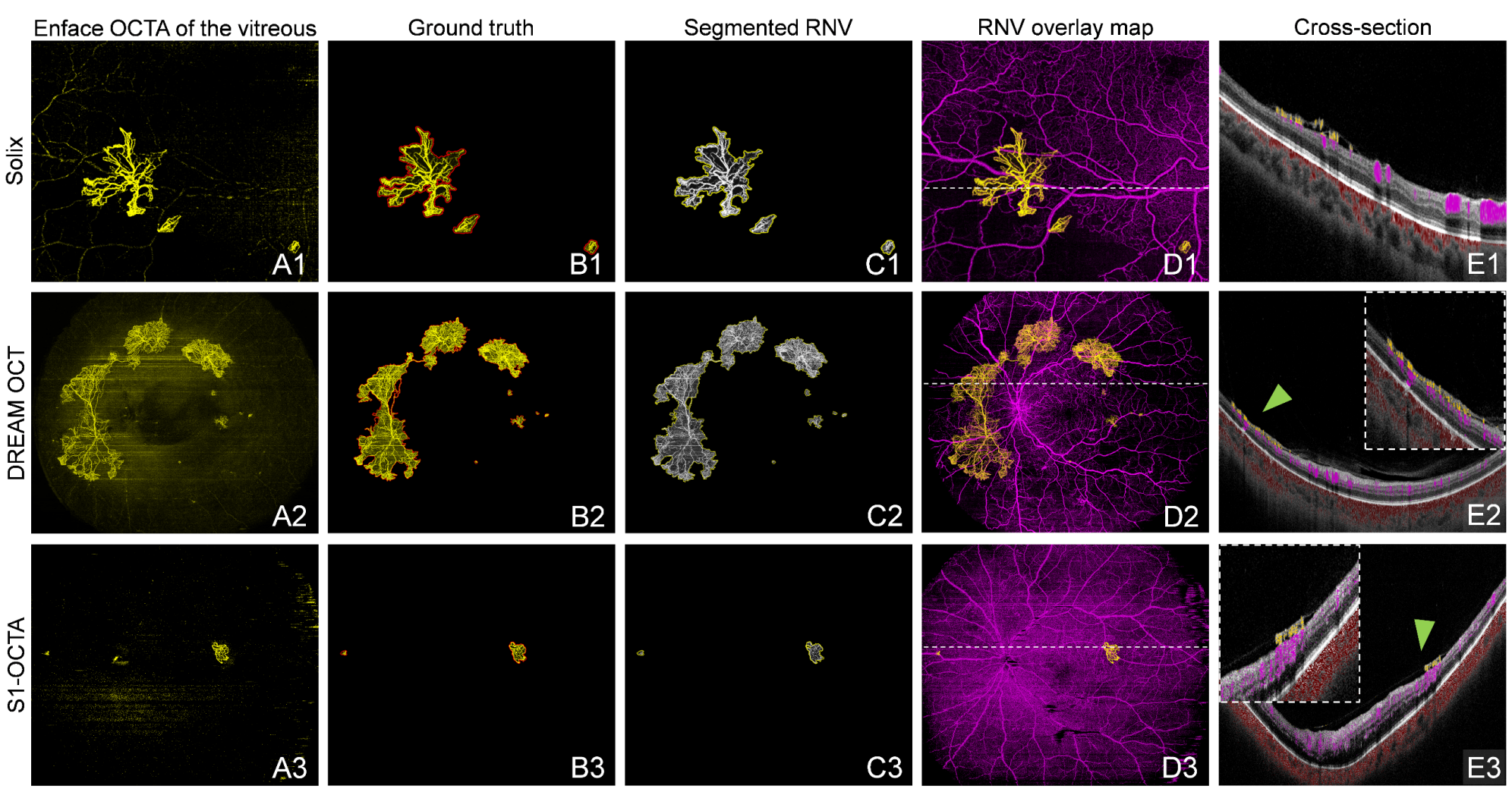

**Fig. 7.** Visualization of RNV segmentation results from three representative cases acquired using different OCTA devices. Each row corresponds to one device: (Row 1): Solix, (Row 2): DREAM OCT, and (Row 3): S1-OCTA. (A) *En face* OCTA projection of the vitreous. (B) Manually annotated OCTA-based ground truth, with the lesion area visualized on *en face* OCTA projection of the vitreous slab. (C) Segmentation results predicted by the proposed algorithm. (D) Segmented RNV overlaid on the original inner *en face* OCTA image. (E) Structural OCT cross-section at the location of the white line in (D), with flow signal overlaid (yellow: RNV flow signals; violet: normal inner retinal vasculature; red: choroidal flow), including blow-up views to highlight the RNV region indicated by the green arrow. The proposed algorithm was able to successfully segment RNV across these three cases from different devices.

### 3.3.3 Performance on clinically challenging conditions

Retinal hemorrhage can sometimes appear as intensely hyperreflective regions on *en face* OCT and may even present with faint OCTA signals, which can lead to misinterpretation as RNV due to their similar structural and blood-flow signal resemblance. These hemorrhagic areas exhibit bright signals on *en face* OCT but lack the well-defined flow patterns characteristic of neovascular lesions (Fig. 8). By jointly leveraging structural and vascular information, the proposed algorithm successfully distinguishes hemorrhage from true RNV, demonstrating robustness in complex pathological scenarios.

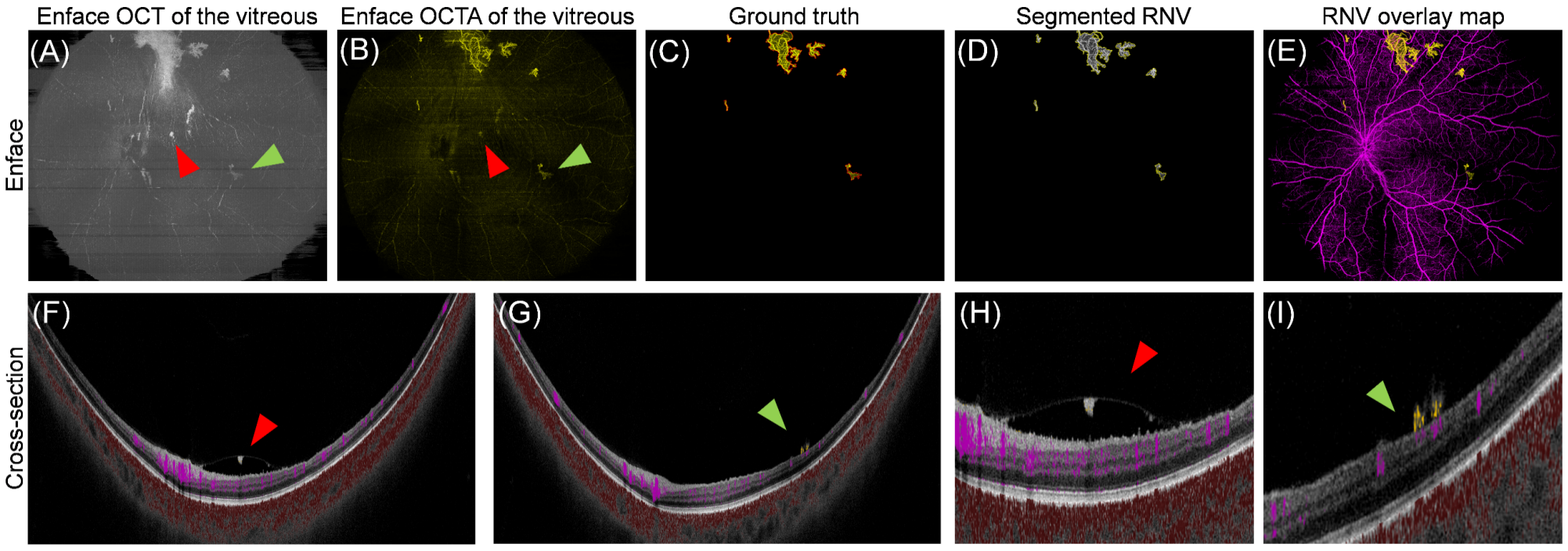

**Fig. 8.** Visualization of RNV segmentation results from a case with hemorrhage acquired using the DREAM OCT device. Throughout, red arrows indicate the location of the hemorrhage age, while green indicates the RNV lesion. (A) *En face* OCT and (B) OCTA projection of the vitreous using the VRI segmentation result. (C) Manually annotated RNV ground truth, with the lesion area visualized on *en face* OCTA projection of the vitreous slab. (D) Predicted segmentation results by the proposed algorithm. (E) Overlay of segmented RNV on the original *en face* OCTA image. (F-I) Structural OCT cross-sectional images with flow signal overlaid (yellow: RNV; violet: retinal; red: choroidal) can help verify algorithm output. (F) Structural OCT B-scan at the hemorrhage site. (G) Structural OCT B-scan at the RNV site. (H) Magnified view of panel F. (I) Magnified view of panel G. The same color scheme applies to all cross-sectional panels (F-I).

### 3.3.4 Performance on challenging scans

Reliable *en face* OCTA projections for RNV analysis are challenged by several types of artifacts (Fig. 9). One of the most problematic artifacts are segmentation artifacts originating from large superficial retinal vessels, which can induce protrusions of the VRI surface. These protrusions often lead to segmentation errors, causing normal retinal vessels to be misclassified as pathological flow. Additional artifacts such as microsaccade artifacts, bulk motion distortion, and decorrelation noise further complicate image interpretation and hinder accurate lesion segmentation [35].

The robustness of the proposed algorithm under suboptimal imaging conditions was evaluated using representative low-quality cases from different devices (Fig. 9). In the DREAM OCT case (Fig. 9 row 1), protrusions of the VRI surface led to segmentation errors that projected superficial vessels into the vitreous slab, creating false-positive signals that affected RNV detection. In another DREAM OCT case (Fig. 9 row 2), weak flow signals and imaging artifacts make it challenging to delineate the RNV membrane even for experienced human graders, yet the algorithm successfully identified the lesion. Similarly, in the S1-OCTA case (Fig. 9 row 3), the lower transverse sampling density in widefield acquisition, together with background noise (due to the ineffective reduction of system jitter noise, eye bulk motion artifacts, etc), may contribute to reduced image contrast and ambiguous boundaries. These factors posed difficulties for both graders and the model, contributing to the lower IOU scores observed on this device. Despite these limitations, the algorithm was still able to localize the RNV membrane, demonstrating reliable performance even under reduced image quality.

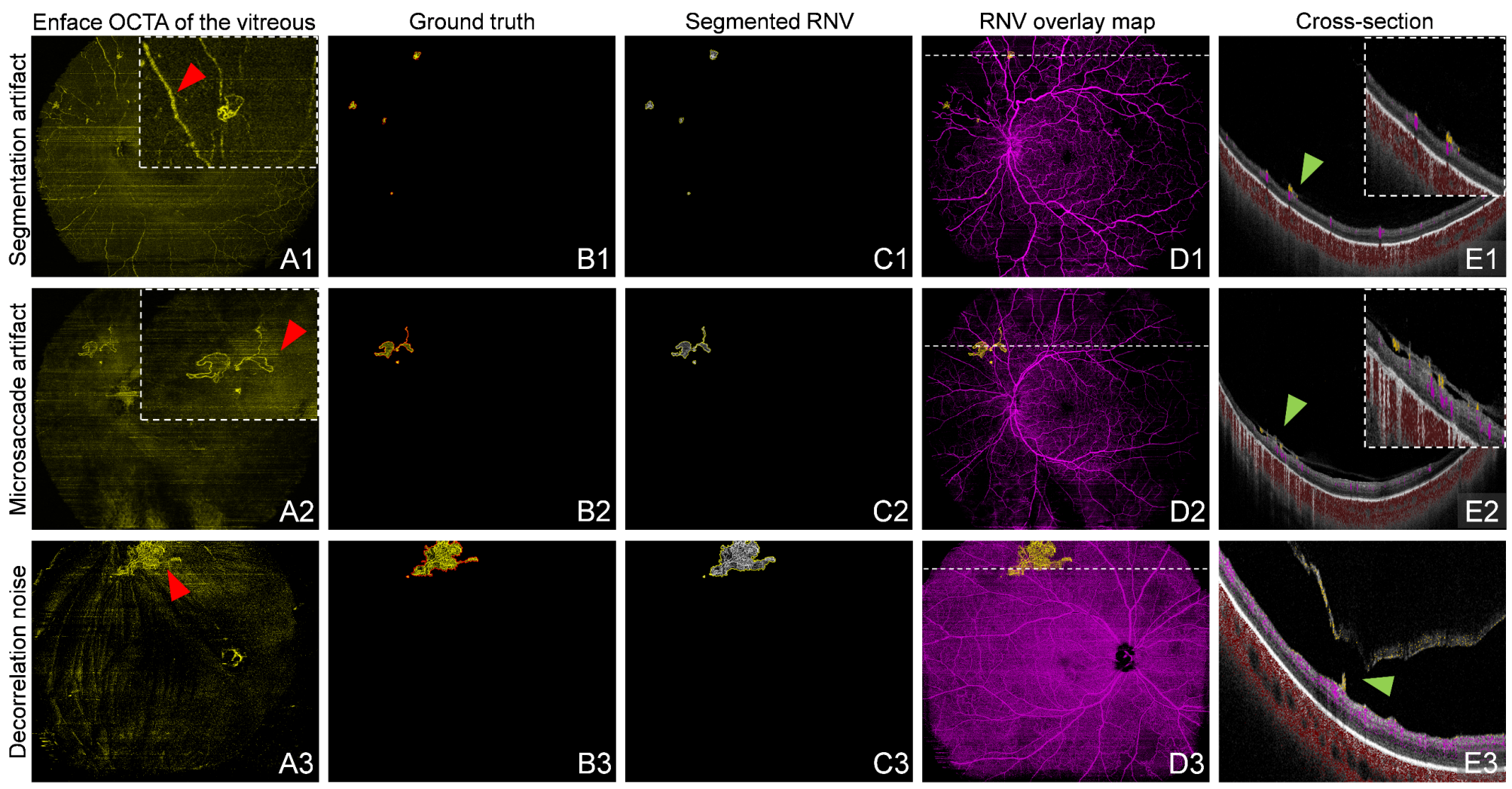

**Fig. 9.** Visualization of RNV segmentation results on low-quality scans from two different devices. (A) *En face* OCTA projection of the vitreous. Red arrows indicate the location of the artifact. (B) Manually annotated RNV ground truth from *en face* OCTA, with the lesion area visualized on *en face* OCTA projection of the vitreous slab. (C) Segmentation results predicted by the proposed algorithm. (D) Overlay of segmented RNV on the original inner *en face* OCTA image. (E) Structural OCT cross-section at the location of the white line in (D), with flow signal overlaid, where yellow indicates RNV flow, violet represents normal retinal vasculature, and red denotes choroidal flow, with blow-up views provided to highlight the RNV region in (E1). Green arrows indicate the location of the RNV lesion. (Row 1): a scan from an DREAM OCT device showing segmentation artifacts from large superficial vessels. (Row 2): a scan from an DREAM OCT device with microsaccade artifacts. (Row 3): a scan from a S1-OCTA device exhibiting pronounced background noise. In all three cases, the algorithm still obtained a results similar to the ground truth.

### 3.3.5 Performance for visualizing ill-defined lesions on fluorescein angiography

FA continues to be a standard clinical procedure for assessing RNV. However, not all RNV lesions that can be identified with OCT and OCTA are captured by FA [9, 36]. Lesions that remain unidentified by FA are often small, meaning that these features could pose a problem for image graders reviewing OCT and OCTA data in high throughput scenarios. We were able to identify cases in which lesions do not appear in FA but were nonetheless correctly segmented using our approach (Fig. 10)

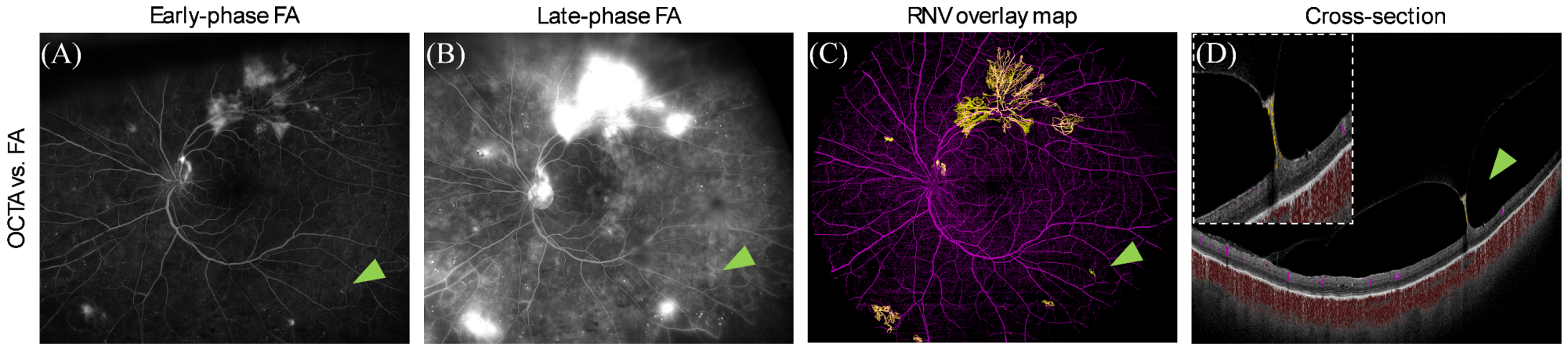

**Fig. 10.** Comparison of RNV detection results between FA and OCTA acquired using the DREAM OCT device. Throughout, green arrows indicate the location of the RNV lesion. (A) Early-phase FA image showing hyperfluorescent neovascular lesions. (B) Late-phase FA image showing leakage from neovascular complexes. (C) Overlay of segmented RNV on the original inner *en face* OCTA image. (D) Structural OCT B-scan at the location of the green arrow in (C) showing flow signals (yellow: RNV flow; violet: inner retinal vasculature; red: choroidal flow), with blow-up views provided to highlight the RNV region.

### 3.3.6 Performance on smaller and larger fields of view

Clinical observations indicate that a substantial proportion of RNV lesions occur in peripheral retinal regions [37]. Consequently, detection frameworks limited to small FOV may fail to capture the full spatial extent of RNV membranes or entirely miss peripheral lesions, underscoring the necessity of wide-field analysis. A notable limitation is the fixed 9 × 9-mm FOV of the Solix device, which restricts the detectable area in OCTA scans. Such constraints often necessitate additional techniques such as image montage to improve FOV and the accuracy of progression tracking (Fig. 11 row 1). In contrast, the DREAM OCT image of the same case (Fig. 11 row 2) demonstrates how a wider scanning area enables full delineation of the RNV membrane, thereby improving the reliability of longitudinal assessments. Our algorithm generalizes well to wide-field OCTA data effectively segments RNV in both of these FOV.

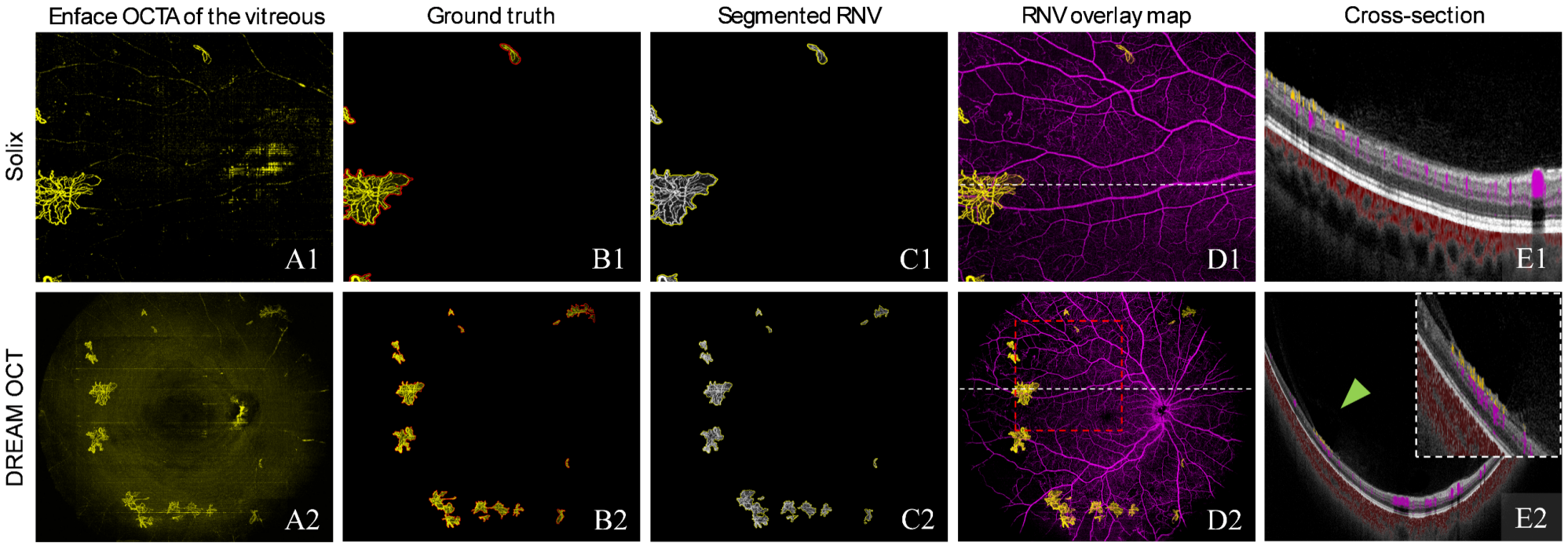

**Fig. 11.** Visualization of RNV segmentation results from a representative case imaged using two different devices. The first row corresponds to the Solix device, and the second row to the DREAM OCT device. (A) *En face* OCTA projection of the vitreous. (B) Manually annotated RNV ground truth on OCTA, with the lesion area visualized on *en face* OCTA projection of the vitreous slab. (C) Segmentation results predicted by the proposed algorithm. (D) Overlay of segmented RNV on the original inner *en face* OCTA image. The red dashed box indicates the region corresponding to the Solix device. (E) Structural OCT cross-section at the location of the white line in (D), showing overlaid flow signals, where yellow indicates RNV flow, violet represents normal inner retinal vasculature, and red denotes choroidal flow, with blow-up views highlighting the RNV region. Green arrows indicate the location of the RNV lesion.

### 3.3.7 Longitudinal monitoring

In addition to detecting RNV lesions, the proposed framework supports segmentation across longitudinal OCT/OCTA scans. Here, for each visit, an *en face* segmentation mask is generated by the model to delineate the neovascular membrane region. Vessel regions are then extracted within the membrane region using a multi-scale adaptive thresholding method combined with morphological operations and constraints, with the goal of enhancing local contrast and suppressing background noise. This refinement helps separate membrane and vessel components across time points. Here, the membrane area is defined as the contiguous region directly predicted by the segmentation model, while the vessel area corresponds to the vessel pixels within the membrane region. As a proof-of-concept demonstration, we applied it to segment RNV membrane areas on 9 × 9-mm OCTA scans from a DR patient with 12 months of longitudinal follow-up. The resulting masks allow quantitative assessment of progression, including changes in lesion area and vascular density over time (Fig. 12).

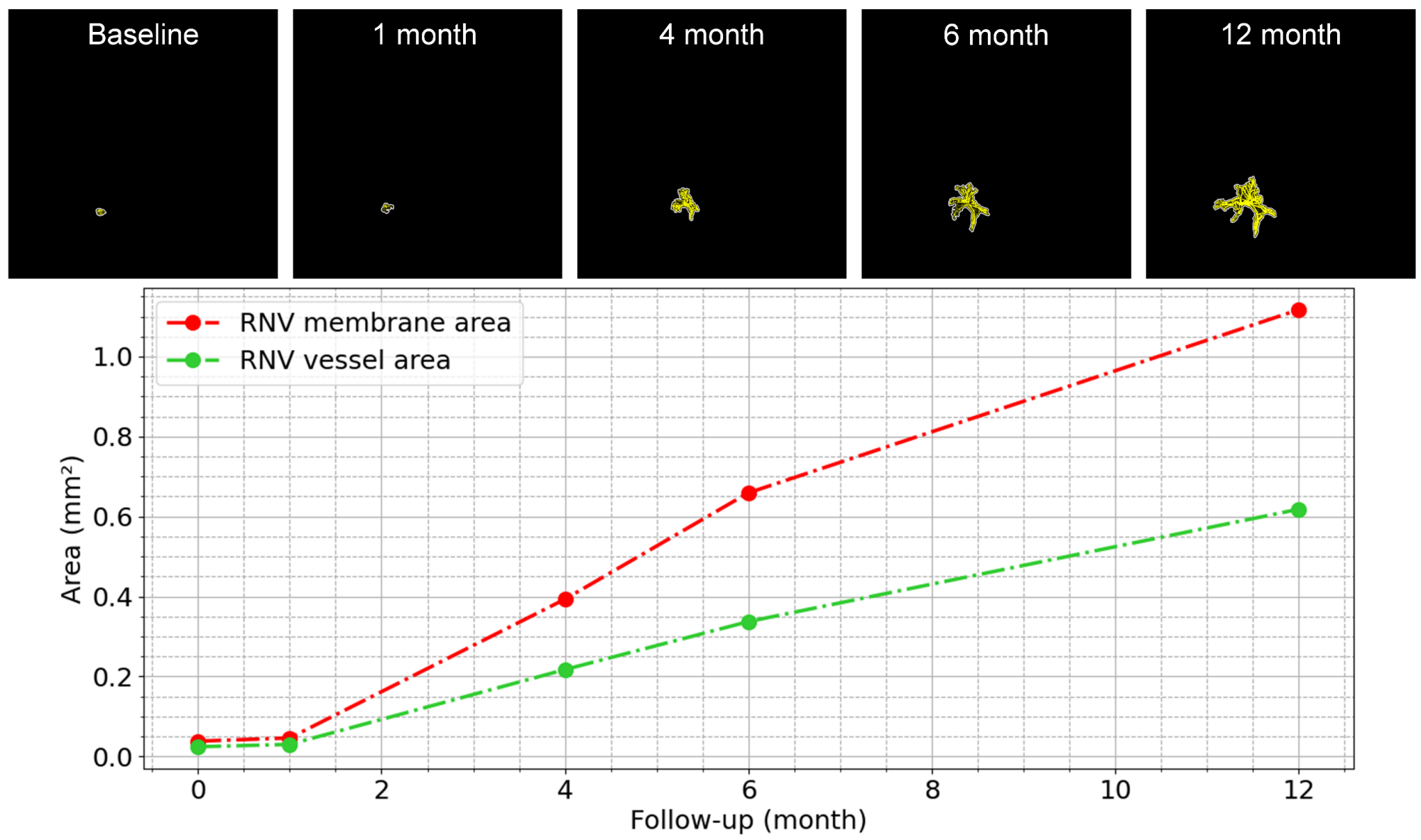

**Fig. 12.** Representative longitudinal segmentation of RNV membranes in a DR patient over 12 months without treatment intervention, shown as a proof-of-concept example of longitudinal monitoring. Each panel shows the segmented RNV region at a specific time point: baseline, 1 month, 4 months, 6 months, and 12 months. Yellow regions denote the refined RNV vessel mask, and white outlines indicate the segmented membrane boundary. Membrane area and vessel area were computed separately at each time point by pixel counting within the corresponding segmentation masks and converted to physical area. The resulting measurements were plotted over time to illustrate longitudinal changes in lesion morphology and vascular components.

## 4. Discussion and conclusion

In this study, we developed a fully automated deep learning framework for the identification and segmentation of RNV in widefield OCT/OCTA data volumes. This framework achieved two primary tasks: (1) diagnosis of RNV, and (2) segmentation of the RNV lesion in *en face* images. The proposed method was evaluated on datasets from two clinics and three different commercial OCTA devices, with no exclusions for poor image quality. Our validation results indicate that our approach is accurate on a clinically realistic dataset. In particular, this work represents an RNV detection approach valid for widefield OCT and OCTA data, which could offer an advantage in clinical practice relative to existing approaches that operate on conventional (smaller) FOV. This performance required several innovations, as indicated below.

First, our approach relied on a novel supporting network that supplied VRI segmentation. Previous studies have commonly employed the VRI slab to detect RNV [13, 38], as it captures extraretinal lesions that extend into the vitreous. However, the effectiveness of this approach is limited by the accuracy of automated segmentation of both the VRI and posterior vitreous membrane. Posterior vitreous membrane segmentation methods often use fixed-height columns to approximate membrane location [14], which can fail to capture the full extent of RNV lesions. Furthermore, previously published VRI segmentation results have been reported to suffer performance decline when assessing flat-type RNV cases [39]. Consequently, in such methods manual correction may be required for accurate detection in such approaches. Our approach addresses this difficulty by introducing two task-specific supporting slabs: a vitreous slab above and a retinal slab below the VRI. Unlike conventional multi-layer retinal segmentation frameworks [40], we reformulate VRI segmentation as a binary localization task. This not only simplifies the model design but also enhances robustness across devices and field curvatures. Notably, the model performs reliably even when neovascular tissue is closely adherent to the VRI. And, since we do not use fixed height columns, we do not risk neovascular vessels extending upward beyond the region of interest defined by such columns. This anatomically informed strategy enables consistent delineation of RNV membranes with irregular morphology or variable height.

Another issue that can deleteriously affect RNV segmentation is the presence of normal large superficial vessels. Segmentation artifacts from large superficial retinal vessels can mimic or obscure true RNV signals in the vitreous, leading to false positives and complicating automated analysis. On *en face* projections, these vessels often produce VRI protrusions that resemble neovascular tufts. These features are particularly challenging to distinguish from true pathology without anatomical context. To address this, we incorporated additional imaging inputs: *en face* OCTA projections of the GCC slab as spatial priors for superficial vasculature, and a vessel-suppressed vitreous OCTA image obtained by subtracting scaled GCC flow signal from the vitreous slab. This strategy, validated in prior studies [25], effectively suppresses segmentation artifacts and enhances the visibility of neovascular tufts, thereby improving the specificity and robustness of RNV detection.

Back to 2015, Jia and colleagues [7, 41] have first defined and demonstrated the way to visualize and quantify RNV in vitreous slab. Other studies have also leveraged anatomical cues for RNV detection. Wang et al [34]. Developed a deep learning system for NVD using 6 × 6-mm OCTA scans, in which optic disc boundary segmentation was used to guide lesion identification within volumetric OCT/OCTA data, achieving strong diagnostic performance within a limited FOV. In contrast, some approaches omit VRI segmentation and instead rely on vessel-derived features. For instance, Tun et al. [12] proposed a feature-based method that computes vessel density, bifurcation density, and vessel thinness from *en face* OCTA scans (12 × 12-mm FOV) of the superficial retinal layer acquired using a swept-source system, followed by support vector machine-based RNV localization. Such methods are primarily designed for coarse localization within small scan areas rather than pixel-level lesion segmentation. By contrast, widefield OCT/OCTA scans exhibit increased retinal curvature and anatomic variation from the posterior pole to the peripheral retina, making direct RNV detection on B-scans more challenging. To address this, our approach incorporates VRI segmentation as a structural prior to construct the VRI-referenced *en face* representations. This provides reliable anatomical guidance for RNV segmentation across widefield retinal images.

Due to these technical considerations our method was successful in diagnosing and segmenting RNV in widefield OCT/OCTA. The transition from conventional FOV to widefield OCT/OCTA has important clinical consequences, particularly in screening for pathologic features that could be found throughout the retina. RNV itself is an example of such a pathology, and in fact it may be more prevalent in the periphery, outside the FOV of macula-centered conventional OCT/OCTA. However, alternative screening approaches like widefield color fundus photography may not provide detection sensitivity commensurate with that provided by OCT/OCTA. Widefield OCT/OCTA combines high detection sensitivity with expansive coverage of the retina. In combination, OCT/OCTA imaging and automated RNV detection as described in this study could provide an improved means of diagnosing RNV. Our approach is also able to assess RNV longitudinal dynamics (Fig. 12), which could provide indications for treatment interventions. In total, we believe that our results demonstrate automated RNV analysis in widefield OCT/OCTA is a promising avenue for ophthalmic translational research.

This study has several limitations. First, the number of eyes without RNV in the dataset is relatively small, which may affect the reliability of eye-level performance metrics. Future studies incorporating a larger number of negative cases will be important to further validate the diagnostic performance of the proposed framework. Second, RNV lesions with blurred or ambiguous boundaries, especially those closely apposed to the VRI, remain difficult to segment accurately. These cases are inherently challenging, even for experienced human graders.

Finally, while the proposed method demonstrates robust performance across most RNV morphologies, segmentation may fail in regions where the VRI is significantly distorted or elevated, or where it is torn, often due to traction exerted by adjacent neovascular tissue. In such cases, manual correction may be required to ensure accurate membrane delineation.

Domain shift across OCT/OCTA systems can lead to variability in model performance. The proposed method was evaluated on data from three different devices in this study. Several design choices were incorporated to address cross-device variability, including intensity normalization, a two-stage framework that transforms volumetric OCT/OCTA data into *en face* representations, and a dual-branch encoding strategy that processes OCT and OCTA modalities separately before feature fusion. While these design choices are intended to mitigate certain sources variation in performance due to domain shift [42], explicit domain adaptation or generalization strategies remain important directions for future investigation.

Building upon these observations, future work will focus on expanding the dataset scale and diversity and exploring the integration of multimodal clinical data such as FA and patient records to provide richer diagnostic context. To further alleviate annotation burdens, we also aim to incorporate semi-supervised or unsupervised learning strategies for VRI and RNV segmentation. These approaches could leverage large quantities of unlabeled OCTA data to improve model robustness while reducing dependence on expert-annotated ground truth. Finally, we will continue refining the model architecture to achieve more accurate delineation of complex neovascular patterns, thereby facilitating broader clinical adoption and real-world scalability.

**Funding.** National Institutes of Health (R01 EY 036429, R01 EY035410, R01 EY024544, R01 EY027833, R01 EY031394, R43EY036781, P30 EY010572, T32 EY023211, UL1TR002369); the Jennie P. Weeks Endowed Fund; the Malcolm M. Marquis, MD Endowed Fund for Innovation; Unrestricted Departmental Funding Grant and Dr. H. James and Carole Free Catalyst Award from Research to Prevent Blindness (New York, NY); Edward N. & Della L. Thome Memorial Foundation Award, and the Bright Focus Foundation (G2020168, M20230081).

**Acknowledgment.** The authors thank Min Gao for her assistance with dataset organization and preparation.

**Disclosures.** Yali Jia: Optovue/Visionix (P, R), Genentech/Roche (P, R, F), Ifocus Imaging (I, P), Optos (P), Boeringer Ingelheim (C), Kugler (R); Tristan T. Hormel: Ifocus Imaging (I); Jie Wang: Optovue/Visionix (P, R), Genentech/Roche (P, R), Ifocus Imaging (I, P); Yukun Guo: Optovue/Visionix (P), Genentech/Roche (P, R); Ifocus Imaging (P); Steven T. Bailey: Optovue/Visionix (F); Kotaro Tsuboi: Alcon Japan (F); Thomas S. Hwang: None;

**Data Availability.** The data sets generated and analyzed in this study are not publicly available at this time but may be obtained from the authors upon reasonable request.